# DDoS Attack and Interception Resistance IP Fast Hopping Based Protocol

Vladimir Krylov and Kirill Kravtsov
Nizhny Novgorod State Technical University n.a. R.E. Alekseev,
Nizhny Novgorod, 603950, Russia
vkrylov@heterarchica.com, kirill@kravtsov.biz

## Abstract

Denial-of-Service attacks continue to be a serious problem for the Internet community despite the fact that a large number of defense approaches has been proposed by the research community. In this paper we introduce IP Fast Hopping, easily deployable and effective network-layer architecture against DDoS attacks. Our approach also provides an easy way for clients to hide content and destination server of theirs communication sessions. We describe a method of dynamic server IP address change and all modules necessary to implement the approach.

## 1 Introduction

A Denial-of-Service attack is characterizes by an explicit attempt to prevent the legitimate use of a service. A Distributed Denial-of-Service attack deploys multiple attacking entities to attain this goal [10]. In such attacks, a single bot or a group of bots are sending a large number of packets that lead to exhausting of victim's bandwidth capacity or software processing capabilities.

According to [10], methods of DDoS attacks can be divided by the two groups: semantic attacks and brute-force (flood) attacks. A semantic attack exploits a specific feature or implementation bug of some protocol or application installed at the victim in order to consume excess amounts of its resources. For example, an attacker can send a specific consequence of packets initiating CPU time consuming procedures on the server. In case of a large number of such requests, the victim is unable to handle requests from legitimate clients. Undesirable impact from such attack can be minimized by protocol or software modifying and by applying of special filter mechanisms. In our paper, we introduce a DDoS defense mechanism that aims to filter all TCP traffic issued by unauthorized clients on network level. Therefore unauthorized malefactor is unable to perform semantic attacks based on TCP protocol on the victim server.

A brute-force attack is an attack aimed to prevent legitimate service using by exhausting of bandwidth. E.g. it is a large number of short requests to the victim which initiates heavy responses sent by the victim. Together these streams overfills bandwidth of the victim server or it's ISP. In contrast to semantic attacks, brute-force attacks abuses legal services so installing of filtering mechanisms for such requests will impact traffic from legitimate client too.

During brute-force DDoS attacks a number of malefactor terminals (botnet) and legitimate users are connected to the victim at the same time (see Figure 1). Each bot sends a big number of requests to the victim which creates heavy malicious traffic targeted at the server. Since the increase in the flow of requests is created here increase the number of terminals, then whichever level of server performance has not been achieved starting from a certain number of bots, they create the flow of requests exceeds the permissible level for any server.

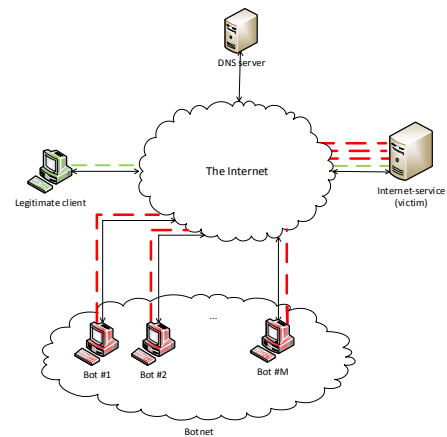

Figure 1: Schema of brute-force DDoS attack

Size and frequency of DDoS attack is continue to grow despite on the fact that a large number of defense mechanisms have been proposed. According to [14] application layer attacks rose approximately 42% in 2013 from 2012, infrastructure layer attacks increased approximately 30% at the same period. The infrastructure layer DDoS attacks are still most popular: around 77% of DDoS attacks in Q4 2013 were infrastructure layer attacks. The average size of attacks increased by 19.5% from Q1 2012 to Q1 2013 (up to 1.77 GB/s) [7]. So, developing new DDoS prevention mechanisms is still a topical issue.

Each new DDoS defense method should satisfy the following main principles:

- Real world applicability. Now a number of different approaches have been suggested in literature which require a significant changes of existing network architecture of ISPs or entire Internet architecture. But the main problem here that mostly DDoS attacks threaten organizations which provide services to end users [13], and so this problem is not very vital for transit ISPs because they actually suffer very little from such attacks. Thus, priority will be given to such systems, which filters malicious traffic without changing of global network architecture, into server's or edge ISP's networks.
- The solution must be designed to prevent misuse. So, it must be not possible to exploit the method to increase impact issues by an attack or to filter legitimate traffic.

## 2  Related work

The rise of DDoS attack frequency in recent years has resulted in many proposed defense approaches from the community. For example, patent [3] suggests solution on application level. According to the article, the server sends a special response on first connection attempt from a client. The client uses this response to identify new URL address, after that the client creates new requests and sends this new requests to this new URL. After receiving this second request, the server validates the new URL based on sent response. If the value does not exceed preset load threshold, the packet will be prioritized and processed by the server. Thus the solution is based on applying of special filter on server side; this filter controls prioritizing of client's requests depending on load on the server at the moment of receiving of the first request from the client.

Other high level approach was suggested under [1]. This method introduced special server responsible for creating and updating of cryptographically secured keys. Each client can access to the service only after successful legitimacy verification on this special server. Thus, client should be successfully authorized on the special server to get a secure key which should be used for processing of a special scenario. The aim of this scenario is identifying client's legitimacy. These approaches purpose defense methods on application layer and does not impact IP packets exchange. So, a malefactor can perform brute force attack on the server.

Also, the research community suggested a wide scope of different more low level approaches. For example, [9] purposes dividing of data stream transmitted between server and client into two consecutive segments on TCP level. This work suggested comparing of keys of two consecutive segments to detect possible segments from not legitimate source. In case of detection of such segments, data receiving will be blocked to prevent possible impact from attack.

Paper [11] introduces DDoS defense mechanism based on dynamic change of server's IP address. Server's IP address is changing according to pseudo-random law which is known only for authorized clients. At the first sight the work [11] purposes a similar DDoS prevention mechanism (dynamic changing of IP address), but this contains some significant differences. Among others, are:

- IP address of the victim is changing only during active DDoS attack on the server
- The new IP address is assigned for all client sessions simultaneously on a relatively long time (suggested period is around 5 minutes)
- Accurate time synchronization is required for calculation of each next IP address since external timestamp is using.

## 3  IP Fast Hopping method

In our paper, we introduce a DDoS prevention mechanism based on protocol level defense methods which was suggested under patent [8]. The main goal of this technique is counteraction to exhausting of server's resources initiated by attackers and prevention of legitimate traffic filtering. To archive these aims, the method is using real-time changing of server's IP address according to a schedule which is available only for authorized clients. Attackers can't get access to this schedule, so they cannot send requests to the correct IP address. Due to this effect, bots are unable to create enough high load on the server to prevent normal system behavior.

In our work, we suggest to call this method as IP Fast Hopping.

The method suggested in this paper is similar to radio systems with frequency hopping. In such systems, receiver and transmitter are switching from one frequency to other frequency synchronously during an ongoing data transmission session. A malefactor's transmitter, which is going to introduce a noise into such session, has not an actual schedule of frequency hopping; therefore such attacker cannot create a noticeable harm for the legitimate transmitter defended by frequency hopping mechanism.

In our case, frequency can be treated like IP address. So, the legitimate client must know schedule of server's IP address changing. At the same time, the schedule should be not available for non-legitimate clients.

The method of IP masquerading for received packets is utilized in Network Address Translation technology. In contrast to the technique suggested in this paper, such IP masquerading is permanent during the entire session, i.e. mapping of the internal constant address to a temporary external address is not changing during a session [5]. This approach provides a way to share limited external network resources between a large number of devices. In our paper, we propose to make such mapping dynamic.

According to DDoS prevention mechanism based on IP hopping approach, DNS entries are equal to IP address of the authorization server instead of IP address of the protected server. To access the protected server, each client must be tested on legitimacy on this authorization server. Authorization process can cover validation of user's login/password, user's subscription on a service and so on. In case of successful client's authorization, the client is redirecting to special server, IP Hopper Manager, instead of to the protected server. This server is controller of enhanced secured sessions. In this paper, enhanced secured session is a communication session between client and server which is protected by IP Fast Hopping method. The legitimate client must establish secured connection to this controller. The IP Hopper Manager will use this connection to transmit a pool of IP addresses and unique identifier of the session to the client's terminal. The IP Hopper Manager sends the same information to a set of edge switchers randomly located in the Internet. These switchers must support IP Fast Hopping method and the server must be signed on this service. In this paper, the edge switcher is high performance switcher which is edge relatively to the suggested protection mechanism, because in the network sector between this switcher and the protected server the data stream is not have any difference in comparison with the case when IP Fast Hopping was not deployed. The IP pool is not ordered and each IP address must be related to this switchers set. Also, this pool should not contain the real server's IP and the "initial" IP. The initial IP is public virtual address of the protected server. All client's applications use the initial IP address instead of real IP address of the server.

After such handshake the client starts communication session with the server. During communication session between the client and the protected server, IP address of the server is hopping between addresses from this pool in real-time. The client's terminal is changing initial IP address in the destination address field of each outgoing packet on an address from the pool of IP addresses according to a special hash function. This hash function is mapping timestamps field of TCP header [6] and unique identifier of the session to an entry of the IP pool. This UID can be obtained for the private key of a certificate installed on the client's equipment or can be received from IP Hopper Manager as was mentioned above. After such replacement of the initial address to a virtual address from the IP pool, the packet is transmitted over the Internet to one of edge switchers according to common switching protocols.

When the edge switcher received the packet from the client, the switcher calculates the same pseudo-random function with the same arguments as was done on the client side. If the result of this calculation is the same to the destination address field of IP header, the packet is forwarded to the real IP address of the server as legitimate packet. Otherwise, this packet will be dropped as malicious packet.

The same procedures (but in reverse mode) will be applied for each server's responses to the client. After receiving of such packets, client's terminal changes server's virtual IP into source field on initial IP address, after that the packet can be processed by client's application by a common way.

As the result, from the point of view of an external observer of the client-server communication session, the IP address of the server is changing regularly to a random address with each increment of timestamps field into TCP header of the packet (usually every millisecond).

Prediction of destination IP of the next packet is very difficult for an external observer due to the fact that destination IP is changing according to pseudo-random function and this observer has not information about parameters of this function (UID or real server's IP address).

If the IP pool is not large enough, a botnet can start an attack on each IP address using masquerading of malicious traffic as legitimate data stream by IP spoofing technique [2]. In this case, edge routers redirects part of hateful traffic together with legal traffic to the protected server. In this paper, we suggest the following options to mitigate such risks:

1. The IP pool which is used for IP Fast Hopping should be large enough to make such excessive attack very resource consuming and non-efficient for possible attackers. Obviously, the method will be more efficient in IPv6 systems. In this case, IP pool can contain a thousands of addresses related to a number of different routers in the Internet.
2. IP providers should apply IP spoofing filtering mechanisms, e.g. [4]

The described particular qualities of introduced DDoS protection mechanism allow to use this method not only for DDoS prevention but also for defense of communication session between server and client. In this paper, all TCP packets in each client-server communication session transmitted via network according to IP Fast Hopping rules without depending on existing of active attack on the server. This fact causes the following effect: for an external observer close to the client, the communication session between the client and the Internet service does not look like packet stream between terminal of the client and a server on which this Internet server is hosted. This session is visible for an external observer as different communication sessions between the client and a large scope of different servers in the Internet and data stream is randomly mixed between these streams. From an external observer point of view, interpretation of these data stream into one logical data stream are difficult process. Also, due to the fact that one pool of virtual IP addresses is shared between different Internet services at the same time, such external observer close to the client is unable to identify server which established communication session. So, IP Fast Hopping could be used in cases when clients want to hide content of data stream and destination of this stream.

## 4 System architecture

Our paper introduces DDoS protection mechanism aims to prevent access to the server from a botnet by dynamic changing of IP address of the server. Such system can have a scope of different implementations, but our work takes into account the following requirement: the suggested approach should be easy deployable in real world conditions and should not require a significant changes of existing network architecture or network equipment. Easy deployability means that switching to the suggested defense method does not require a considerable preparation or workflow changes for an Internet server or its clients.

To archive these goals, our work uses only existing commonly used technologies and protocols and also the logical core of the system is deployable into external (for the server and its clients) networks (e.g. into ISP networks).

We can say that our work is a new point of view on using of existing abilities of TCP/IP protocols. Our paper introduced re-use of already used technologies for DDoS attacks prevention. An example of such alternative utilization, the timestamps field of TCP packet header, are suggested to be used not only to identify the correct packets order [6], but at the same time this field can be used to identify the correct destination address of the packet as was described above.

Also, our work introduced distributed DDoS defense mechanism: an Internet server is being hidden behind a large pool of virtual IP addresses which belong to a big number of routers in different sectors of the Internet. Since this IP address pool is public, botnets can initiate DDoS attacks on one or several of these IP addresses. But the pool is divided into groups of addresses which belong to various routers in various Internet sectors. So the stream of malicious packets initiated by a botnet is divided into several sub-streams directed to several Internet sectors by commonly used switching protocols. And, according to our work, this stream will be filtered into this different networks. This approach defends the victim server and also our method decreases load on network infrastructure of victim and it's ISP during active DDoS attack.

In case of deploying of the introduced defense mechanism, the original client-server architecture (see Figure 1) contains some new blocks (see updated schema on Figure 2). The suggested architecture has the following difference:

1. As was mentioned above, the DNS server contains link to IP address of Authorization server instead of IP address of Internet-service

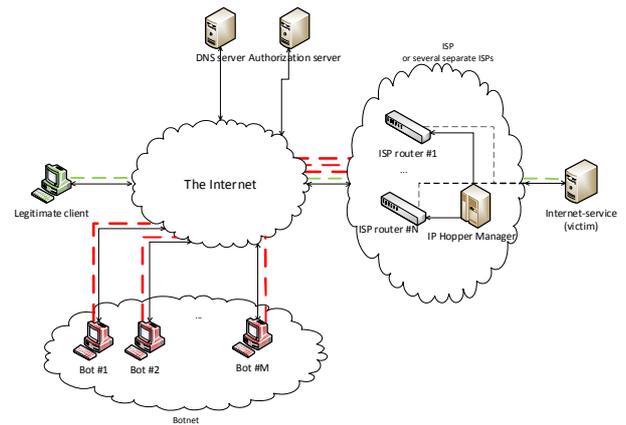

Figure 2: IP Fast Hopping architecture

2. Introduced Authorization server which validates legitimation of the client. If client successfully authorized and the client's terminal has a special SSL certificate and supports IP Fast Hopping algorithm (i.e. installed special software – IP Hopper Core), Authorization server initiates handshake between the client's IP Hopper Core and IP Hopper Manager
3. IP Hopper Core is special system utility installed on client's terminal and ISP routers #1 - #N (entire IP pool belongs to these routers). This utility is performing establishing of enhanced secured connection and real-time changing of initial IP address of the Internet-service on one address from IP pool according to rules of IP Fast Hopping.
4. IP Hopper Manager is a server which is responsible for controlling the enhanced secure connections between Internet-services and clients.

Time chart of introduced defense mechanism can be found on Figure 3.

## 5 Implementation

As was noted above, one of requirements for our work is real world deployability. Therefore, we implement IP Fast Hopping mechanism as kernel module of OS GNU/Linux. In this case, installing this module on routers based on GNU/Linux is enough to deploy the suggested system. In our work we build such routers based on Debian OS.

Linux kernel contains built-in firewall Netfilter [12], which is responsible on packet filtering and forwarding according to predefined rules by iptables utility. Netfilter architecture is scope of hooks of ordered rules. Netfilter performs a predefined action with a packet, which is passed to a hook, according to the corresponding rule.

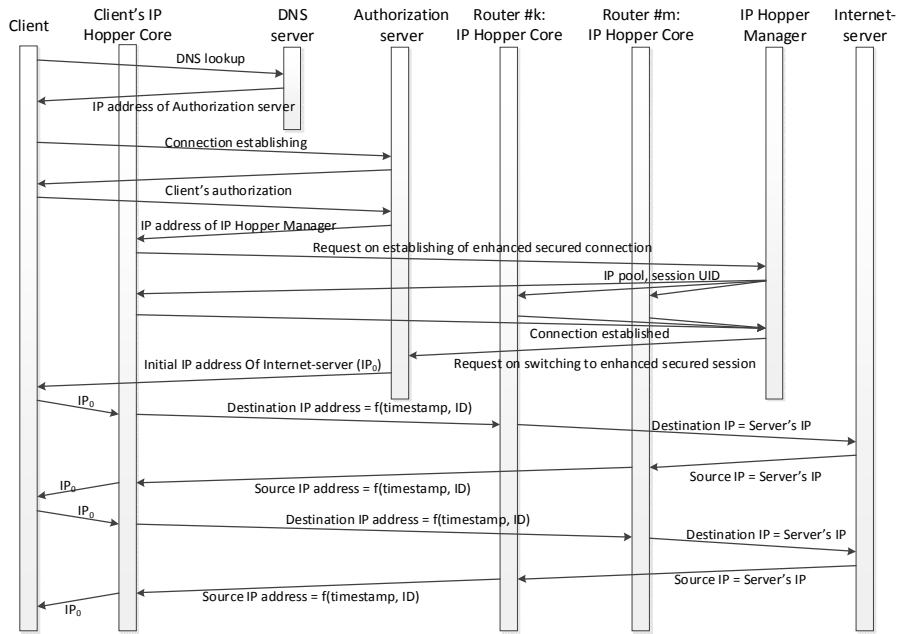

Figure 3: Time Chart of IP Fast Hopping

Netfilter supports 5 hooks: PREROUTING, INPUT, FORWARD, OUTPUT, POSTROUTING. When the packet comes to the system, the packet is processed by PREROUTING hook. If this packet is addressed to a local process, it is passed to INPUT hook, otherwise it is passed to FORWARD. All packets sent by local processes are processed by OUTPUT hook. The final processing of the packet outgoing from the system (forwarded under FORWARD hook or issued by a local process) is performing by POSTROUTING hook.

In our work, Netfilter contains new module which is responsible for changing of IP address into destination field of outgoing packets and into source field of ingoing packets. This module is calculating the new IP address according to IP Fast Hopping rules (by timestamp field and session UID). During handshake, IP Hopper Manager adds new set of rules into POSTROUTING hook on client's terminal and into PREROUTING each edge switcher. This rule activates the kernel module which implements the following algorithm:

- On the client side this module calculates hash-function using timestamps field and session UID for each outgoing packet addressed to the initial IP address. After that the module uses this result as index of correct address into IP pool which should be put into destination field of the packet. For each ingoing packet from the same communication session, the module performs the same actions for source field: checks the current value of the field (by calculation of the same hash-function) and changes it on the initial address.

- On switchers side this module calculates hash-function using timestamps field and session UID for each ingoing packet addressed to IP addresses from IP pool. If the current destination address corresponds to the timestamps field and session UID, the real IP address of the server will be placed into the destination field. Otherwise, the packet will be dropped. For all ingoing packets issued by the server, the module will replace source field by one of virtual addresses according to current value of hash-function.

## 6  Conclusions

We presented IP Fast Hopping, a new approach that can prevent exhausting of server's resources during brute-force DDoS attacks and can be used to hide content and destination of client's communication session. This method hides the real IP address of the server behind a big number of "virtual" IP addresses. The mapping of the real IP address on one of "virtual" is unique for each communication session and changes dynamically every millisecond. The introduced approach is distributed: it divides the traffic from legitimate users and botnets into a number of sub-streams. This leads to a decrease of load on network infrastructure during active DDoS attack. The method is easily deployable and can filter even the biggest malicious streams.